\documentclass[12pt,preprint]{aastex}
\newcommand\bl{{\mathbf l}} 
\newcommand\bm{{\mathbf m}}

\newcommand\bs{{\mathbf s}}

\newcommand\bB{{\mathbf B}}

\newcommand\bP{{\mathbf P}}
\newcommand\bQ{{\mathbf Q}}

\newcommand\rmd{\mathrm{d}}
\newcommand\rmi{\mathrm{i}}
\newcommand\rme{\mathrm{e}}

\newcommand\norm[1]{\lVert #1 \rVert}

\usepackage{amsmath,amsfonts}
\begin{document}
\title{The visual orbits of the spectroscopic binaries HD 6118 and 
HD 27483 from the Palomar Testbed Interferometer}
\author{Maciej Konacki \altaffilmark{1}}
\affil{Department of Geological and Planetary Sciences, California
Institute of Technology, MS 150-21, Pasadena, CA 91125, USA\\
Nicolaus Copernicus Astronomical Center, Polish Academy of Sciences,
Rabia\'nska 8, 87-100 Toru\'n, Poland}
\author{Benjamin F. Lane \altaffilmark{2}}
\affil{Center for Space Research, MIT Department of Physics,
70 Vassar Street, Cambridge, MA 02139, USA\\
Department of Geological and Planetary Sciences, California
Institute of Technology, MS 150-21, Pasadena, CA 91125, USA}
\altaffiltext{1}{e-mail: maciej@gps.caltech.edu}
\altaffiltext{2}{e-mail: blane@mit.edu}
\begin{abstract}
We present optical interferometric observations of two double-lined 
spectroscopic binaries, HD~6118 and HD~27483, taken with the Palomar 
Testbed Interferometer (PTI) in the K band. HD~6118 is one of the most 
eccentric spectroscopic binaries and HD 27483 a spectroscopic binary in 
the Hyades open cluster. The data collected with PTI in 2002-2003 allow 
us to determine astrometric orbits and when combined with 
the radial velocity measurements derive all physical parameters of the 
systems. The masses of the components are $2.65\pm0.27 M_{\odot}$ 
and $2.36\pm0.24 M_{\odot}$ for HD~6118 and $1.38\pm0.13 M_{\odot}$ and 
$1.39\pm0.13 M_{\odot}$ for HD~27483. The apparent semi-major
axis of HD~27483 is only 1.2 mas making it the closest binary successfully
observed with an optical interferometer.
\end{abstract}

\section{Introduction}

Following Michelson's suggestion and his success in measuring diameters
of Jupiter satellites with an optical interferometer \citep{Mich:1891::}, 
Schwarzschild managed to resolve a number of double stars
with his own grating interferometer \citep{Sch:1896::}. Not long
afterward, Anderson determined the first visual orbit of Capella
\citep{And:20::}, later improved by Merrill \citeyearpar{Mer:22::} with the
same instrument --- the rotating interferometer attached to the 
100-inch telescope at the Mount Wilson Observatory and originally 
used by Michelson. In the same experiment, Merrill also 
attempted to resolve a number of spectroscopic binaries but without success. 

The modern generation of interferometers developed in the 80's 
and 90's finally allows to resolve and subsequently determine visual 
orbits of spectroscopic binaries in a fairly routine manner. Up to date
about twenty spectroscopic binaries have had their orbits
determined with optical interferometers \citep[for a review see][]{Qui:01::}. 
The Palomar Testbed Interferometer (PTI) itself has been successfully used to 
determine a number of visual orbits of double-lined spectroscopic binaries 
including RS Cvn \citep{Kor:98::}, $\iota$ Peg \citep{Bod:99a::}, 
64 Psc \citep{Bod:99b::}, 12 Boo \citep{Bod:00a::}, BY Dra \citep{Bod:01::}  
and Gliese 793.1 \citep{Tor:02::}. The other instruments that have been 
successfully used in binaries studies include the Mark-III 
\citep{Sha:88::} and the Navy Prototype Interferometer \citep{Arm:98::}.

The capability to determine visual orbits of spectroscopic binaries is an 
instrumental achievement very important for stellar astronomy. It allows
to derive all geometric orbital elements, precise magnitude differences
between the components and their angular diameters. This information
combined with spectroscopic and photometric data yields colors,
luminosities, masses and distance to the system. For binaries with 
resolved components it also provides their radii and temperatures.
It is noteworthy that in some cases the derived component masses are
already at the $1-2$ percent precision level sufficient to constraint modern main 
and post-sequence stellar models \citep[see e.g.][]{Tor:02::}. 
Significant progress in this area 
is expected when the new generation of optical interferometers 
offering baselines longer than a hundred of meters, e.g. 
CHARA \citep{McA:00::}, and/or large apertures e.g. the Keck Interferometer 
\citep{Col:00::} or the Very Large Telescope Interferometer \citep{Gli:00::}
become fully operational. Still, observing spectroscopic binaries with 
currently available optical interferometers is a sensible goal as the 
potential of these instruments and the reservoir of accessible targets 
has not been depleted. Interferometric observations
of currently accessible spectroscopic binaries combined with other observables 
deliver a complete set of physical parameters for the components in a much
shorter period of time than astrometric observations of 
visual binaries (which have orbital periods measured
in years) and in contrast to eclipsing binaries can be successfully
carried out for systems with arbitrary orbital inclinations.

In Spring 2001 we have undertaken at the PTI an observing program to
determine the orbits of resolvable spectroscopic binaries from the Batten
catalog \citep{Bat:89::,Bat:97::}. Here we report our first results --- the visual 
orbits of two double-lined binaries, HD 6118 one of the 
most eccentric spectroscopic binaries and HD 27483 a spectroscopic binary in
the Hyades open cluster, which were observed with the PTI in K band 
(2.2 $\mu$m) band in 2001 and 2002.

\section{Interferometric observable}

\subsection{Theory}

From the theoretical standpoint optical interferometry differs little
if at all from its radio counterpart. The fundamental observable from 
a two aperture interferometer is the fringe visibility (also called 
coherence function or just visibility) represented by a (usually complex)
number $\Gamma = V\rme^{-i\phi}$ where $V$ is its amplitude and $\phi$
phase. The relation between visibility and the source intensity
distribution $I(\Delta\alpha,\Delta\delta)$ is given (under source
incoherence and small-field approximations) by the van
Cittert-Zernike Theorem \citep{Bor:99::,Tho:01::}
\begin{equation}
\label{vis:01::}
\Gamma(u,v) = \int\int I(\Delta\alpha,\Delta\delta)
\rme^{-2\pi\rmi(u\Delta\alpha + v\Delta\delta)}\Delta\alpha\Delta\delta,
\end{equation}
where $(\Delta\alpha,\Delta\delta)$ are the coordinates in the tangent plane
of the sky and $(u,v)=\bB_{\perp}/\lambda$ are the coordinates of the projected
baseline vector ($\bB$ represents the spatial separation of the apertures) 
in the same plane and expressed in units of the observing wavelength
$\lambda$. 

The amplitude of the visibility has dimensions of power and corresponds to 
the amount of power that the interferometer measures. Real instruments 
typically produce the normalized (by the total power received from the
source), dimensionless visibility amplitude, $0 \le V \le 1$. For our
further considerations two model visibility amplitudes are particularly
useful: that of a uniform disk (i.e. a single star), and that of a binary system 
comprised of two model disks. It easy to show that these are given by 
respectively \citep[see][]{Bod:00b::}
\begin{equation}
\label{disk:02::}
V^2_{disk} = \left(\frac{2J_1(\pi\theta B_{\perp}\lambda)}
{\pi\theta B_{\perp}\lambda}\right)^2
\end{equation}
where $B_{\perp} = \norm{\bB_{\perp}}$ is the length of the projected
baseline vector and
\begin{equation}
\label{bin:03::}
V^2_{binary} = \frac{V_1^2 + r^2 V_2^2 + 2 r V_1 V_2 
\cos{( 2\pi \bB_{\perp}\cdot\Delta\bs /\lambda)} }{(1 + r)^2}
\end{equation}
where $V_1,V_2$  are the visibilities of individual stars 
given by \eqref{disk:02::}, $r$ is the brightness ratio at the observing
wavelength $\lambda$ ($r = P_2/P_1$ where $P_1,P_2$ are the total powers
of the binary components at the  given wavelength) and 
$\Delta\bs = (\Delta\alpha,\Delta\delta)$ is the 
separation vector between the primary and the
secondary in the plane tangent to the sky.

\subsection{Practice}

The Palomar Testbed Interferometer (PTI) developed by the Jet Propulsion
Laboratory, California Institute of Technology, for NASA to test
interferometric techniques, is a long-baseline infrared interferometer 
located at Palomar Observatory, California. It consists of three
fixed 40 cm apertures that combined pairwise offer baselines between
86 and 110 m. It operates in two wavelength ranges K ($\sim 2.2\mu$m) 
and H ($\sim 1.6\mu$m). Full details of the instrument's architecture
and operation can be found in \cite{Col:99a::}. The data reduction and
calibration procedures for the instrument are described in 
\cite{Bod:98::,Col:99b::,Lan:00::}.

Optical interferometers typically produce only square of visibility 
amplitude, $V^2$. The reason for this is that the phase of the measured visibility 
is usually corrupted by the atmosphere. In addition, we are able to construct 
unbiased $V^2$ estimators \citep{Col:99b::,Moz:91::}. 
In the case of PTI data, the square of visibility amplitude is estimated
as the average of the visibilities from all the channels of a given
band (K, 2.0-2.4 $\mu$m; H, 1.4-1.8 $\mu$m) and corresponds to 
the SNR-weighted mean wavelength, $\lambda$. The remaining step that finally 
results in a reliable $V^2$ for an astronomical source is the calibration process. 

Ideally, an interferometer's response to a point source would be $V^2 = 1$. However 
since generally speaking a real instrument is not perfect and does not operate in 
a perfect environment, the response to a point source, called system visibility, 
is $V_{sys}^2 < 1$ and this value changes temporally (varying seeing index being 
one of the contributors) and spatially (e.g. with apparent source position). Thus 
the calibration process involves interlacing target observations with calibrator 
observations and using $V^2_{sys}$ to calibrate the target observation in a simple 
but effective way \citep{Bod:98::,Moz:91::}
\begin{equation}
\label{pra:1::}
V^2_{calibrated} = V^{2}_{measured}/V^2_{sys}
\end{equation}
Choosing an appropriate set of calibrators is an important part of the 
interferometric observation. The choice of calibrators is dictated by
a few, mostly common sense requirements: (1) a calibrator should have similar
brightness to the target (although it should not be too faint if the
target happens to be faint for a given interferometer)
(2) it should be located in the sky not far from the target
(3) finally, an ideal calibrator should resemble a point source as closely
as possible. The last requirement is dictated by the estimation
of the system visibility through
\begin{equation}
\label{pra:1a::}
V^2_{sys} = V^{2}_{cal-measured}/V^2_{cal-expected}
\end{equation}
where $V^{2}_{cal-measured}$ is a measured visibility and $V^2_{cal-expected}$ 
is an expected (model) visibility of a calibrator. From the above
equation, it is clear that a calibrator should have as simple model visibility
as possible to avoid introducing any model dependent systematic errors
in the estimated system visibility. It can be realized quite well using
single stars. Their model visibility can be described using equation 
\eqref{disk:02::} where the only unknown is the diameter of the star.
With few exceptions, the diameters of stars are not know from direct
observations. Hence to minimize the errors in the estimated
system visibility due to errors in the (estimated) diameters of the 
calibrators (note that $\rmd{V_{sys}}/\rmd{\theta} \rightarrow 0$ as $\theta \rightarrow 0$), 
it is best to use calibrators which are as unresolved 
as possible (or alternatively use stars which have diameters {\it very well} 
determined). For the purpose of the calibration process, one can determine 
angular diameters of the calibrators by fitting the black body model to their 
archival photometric data. Assuming that the calibrators are essentially
unresolved, such a procedure (i.e. errors in the estimated angular
diameters) will not corrupt the calibration process.

It is sufficient to have two calibrators. However since
more than 1/2 of all stars are in binary or multiple systems, one will
sooner or later find out that one of the calibrators is a previously
unknown binary. Hence a set of three calibrators is a practical and 
safe choice. From experience with PTI it follows that  it is best to observe 
targets and calibrators in interleaved, short period ($\sim10$ minutes) intervals
to properly address any sky position-dependent and temporal variations 
in the system visibility. For the purpose of the orbit determination of a 
spectroscopic binary, it is beneficial to observe a target over 2-3 hour
time span and gather $V^2$ measurements corresponding to different projections
of the separation vector $\Delta\bs$ onto the baseline vector $\bB_{\perp}$
(conf. equation \eqref{bin:03::}). Also a final set of $V^2$ measurements 
should correspond to a fairly complete orbital phase coverage. An incomplete 
coverage may cause significant problems in assessing if the best-fit solution 
is indeed the correct orbital solution (see next section).

In the end, the observer is served by the reduction pipeline
with the following six numbers $\{t_i,V^2_i,\sigma_i,\lambda_i,u_i,v_i\}$ where
$t_i$ is the time of observation, $V^2_i$ is the calibrated visibility
amplitude squared, $\sigma_i$ its error, $\lambda_i$ is the mean
wavelength for the observation, and $(u_i,v_i)$ are the components of the
projected baseline vector. While a (very small) error in $t_i$ is 
completely unimportant for further considerations, $\{\lambda_i,u_i,v_i\}$
are known with finite precision and a proper error analysis should take
their uncertainties into account. Also to perform the calibrations,
the diameters of the calibration stars are necessary. Since these 
are known with finite precision, 
their uncertainties should also be taken into account.

\section{Data modeling}

The interferometric observations of spectroscopic binaries 
share the description of the binary motion with regular
astrometry of visual binary stars. The motion of the primary with respect
to the secondary is thus described by the following equations
\cite[see e.g.][]{Kov:95::,Kam:67::}
\begin{equation}
\label{e:RtKepler}
   \Delta\bs(t) = \left(\Delta\alpha(t),\Delta\delta(t)\right) =  
\kappa a\left(\bP\,(\cos E(t) - e) + \bQ\,\sqrt{1 - e^2}\sin E(t)\right),
\end{equation}
\[
\bP = \bl\,\cos\omega + \bm\,\sin\omega,
\qquad
  \bQ = -\bl\,\sin\omega + {\bf
m}\cos\omega,
\]
\[
  \bl = (\cos\Omega, \sin\Omega), \quad
  \bm = (-\cos i\sin\Omega, \cos i\cos\Omega)
\]
where $E=E(t)$ is the eccentric anomaly given by the Kepler equation
$E - e\sin E = 2\pi(t - T_{\mathrm{p}})/P$, $P$ is the orbital period,
$a,e,i,\omega,\Omega,T_p$ are the standard Keplerian
elements---the semi-major axis, the eccentricity, the inclination, the
longitude of pericenter, the longitude of ascending node and the time
of periastron, and $\kappa$ is the parallax. The above equations can 
be used together with the equation \eqref{bin:03::} to model the 
observed visibilities. The typical approach is the least-squares fit 
with respect to the following set of parameters $\{r,\hat{a},P,e,i,\omega,
\Omega,T_p\}$ where $\hat{a} = \kappa a$ is the apparent semi-major
axis. Obtaining the set of parameters that truly correspond to the global
minimum of $\chi^2$ is more tedious than in the case
of radial velocity or astrometric data. From the relative orbit
alone there are a few possible solutions for the pair of angles 
$\omega,\Omega$. Namely $\omega,\Omega$ as well as
$\omega\pm\pi,\Omega\pm\pi$. Also, since the visibility itself depends
on the scalar product $\bB_{\perp}\cdot\Delta\bs$, it is 
invariant under the rotations of the separation vector $\Delta\bs$ by $\pi$.
Effectively, $\chi^2$ has many local well defined minima and in order
to find the global minimum one needs to carry out the least-squares
fit on a grid of initial values of the model parameters that covers
the entire parameter space.

Spectroscopic binaries that are observed with optical interferometers
have usually well determined orbits from radial velocity (RV) measurements.
It substantially reduces the computational effort to obtain the best-fit
parameters as the spectroscopic orbit already supplies most of them
(e.g. it enables to resolve ambiguity in $\omega$ but not in $\Omega$). 
In fact it is useful to combine the fit to radial velocities and
visibilities into one procedure by defining $\chi^2$ as
\begin{equation}
\chi^2 = \sum_{i=1}^N (V_i^2 - \widehat{V}_i^2)^2/\sigma_{V^2_i}^2
+ \sum_{i=1}^N (RV_i - \widehat{RV}_i)^2/\sigma_{RV_i}^2
\end{equation}
where $\sigma_{V^2},\sigma_{RV}$ are the measurement errors and
$\widehat{V}^2, \widehat{RV}$ denote model observables,
and performing the fit over the entire set of parameters including those
intrinsic to the radial velocity model. To perform the
fits, we employ the Levenberg-Marquard algorithm \citep[see e.g.][]{Pre:92::}. 
The formal errors of the fitted parameters can easily be computed within the
least-squares fit formalism. However one cannot forget about the additional
systematic errors that come from several sources (1) the uncertainty in the 
projected baseline $(u,v)$ (2) the uncertainty in the mean
wavelength $\lambda$ (3) the uncertainties in the diameters, $\theta_j$,
of the calibrators (which affect the modeled visibility through equations 
\eqref{pra:1::}) and the components of the binary (which affect
the modeled visibility through equations \eqref{bin:03::}).
For the purpose of this work we adopt that following conservative, based
on our experience with the PTI, estimates of these uncertainties 
(1) 0.01 percent in $(u,v)$, (2) 1 percent
in $\lambda$  and (3) 10 percent in the
calibrator and binary components diameters.

\section{HD 6118}

HD~6118 ($\sigma$ Psc, HR~291, HIP~4889) is a double-lined spectroscopic binary 
with an orbital period of 81.13 days, a spectral type of both components 
of B9.5V, $V = 5.47$ mag, $(B-V)=-0.05$ mag and $(U-B) =-0.18$ mag \citep{Cow:69::}.
Its binary nature was determined by 
\cite{Cam:18::}. The spectroscopic orbit was derived by \cite{Bel:47::} based 
on 28 radial velocity measurements obtained in the years 1944-45 with the Lick 
Observatory Mills 3-prism spectrograph. The average error of these velocities 
is 5 km/s and it is the only set of RVs published for this star. HD~6118 has 
an orbital eccentricity of 0.89. The distance to the star is $127 \pm 11$ pc 
from the Hipparcos parallax measurement of $7.87 \pm 0.68$ mas \citep{Per:97::}.

We observed HD~6118 in 2002 and collected 69 $V^2$ measurements in K band
(see Table 4). Th best-fit solution to archival radial velocity (RV) and our
$V^2$ measurements is shown in Table~2. It is characterized by an rms
of $4.8$~km/s and $3.6$~km/s in the RVs of the primary and secondary
and $0.066$ in $V^2$ (see Fig.~1-2). The reduced $\chi^2$ of the combined
RV/astrometric solution is 1.06. The derived physical parameters of the 
primary and secondary are shown in Table~3. The masses of the components
are reasonably well determined as $2.65\pm0.27 M_{\odot}$ and 
$2.36\pm0.24 M_{\odot}$. The accuracy of the parallax, $8.86\pm0.07$ mas, 
is ten times better than from Hipparcos but consistent with Hipparcos
determination at $1.5$ of its formal error. The masses of the components
accurate at the $10$ percent level do not allow us to perform any challenging
tests of stellar evolution models. Nevertheless, they can be used to estimate
the age of HD~6118. To this end, we used the theoretical isochrones from
\cite{Ber:94::} and the K-band photometric data for HD~6118 from the 2MASS
catalog \citep{Cut:03::}. Even though we do not have any abundance 
information for the star, the error in the age is dominated by the errors in the
component masses. From Figure~4 it follows that the age is 160-200 Myr 
(depending on the assumed abundance) with a rather large error of about 
100-150 Myr. Finally, as can be seen in Figure~3, the apparent separation
between the components is always too large to allow for an eclipse (the 
estimated angular diameters of the components are $\sim0.15$ mas while
the smallest orbital separation is about 0.8 mas).

\section{HD 27483}

HD~27483 (HR~1358, HIP~20284) is a double-lined spectroscopic binary in the Hyades open cluster
It has an orbital period of 3.06 days, a spectral type for both components 
of F6V, $V = 6.16$ mag, $(B-V)=0.46$ mag and $(U-B) =0.02$ mag \citep{Joh:68::}.
The radial velocity orbit was determined by Northcott and Wright 
\citeyearpar{Nor:52::} and improved by Mayor and Mazeh \citeyearpar{May:87::}. 
The distance to the system is $45.9\pm1.8$ pc 
from the Hipparcos parallax measurement of $21.8\pm0.85$ mas. 

We observed HD~27483 in 2001 and 2002 and collected 81 $V^2$ measurements in 
K band (see Table 7). Unfortunately, HD~27483 is only partially resolved with 
PTI because its semi-major axis is only 1.2 mas. In such a case there is 
a strong correlation between the apparent semi-major axis, $\hat{a}$, and the
brightness ratio, $r$. In effect, it is not possible to determine both
parameters reliably \cite[see also][]{Kor:98::}. Therefore, in order to 
derive the inclination of the system we proceeded as follows. One can assume 
that the brightness ratio is close to 1, since the mass ratio of both stars is very close
to 1. Also, one can use the parallax measurement from Hipparcos, $\kappa$ , to 
additionally constrain the least-squares fit (through $\hat{a} = a\kappa$). 
With such a setup ($r=1$, $\kappa=21.8$ mas), we fitted for the remaining 
orbital elements. The best-fit solution (see Table~5) to the archival radial 
velocity (RV) from \cite{May:87::} and our $V^2$ measurements is characterized 
by the rms of $1.1$~km/s and $1.8$~km/s in the RVs of the primary and secondary
and $0.045$ in $V^2$ (see Fig.~5-6). The reduced $\chi^2$ of the combined
RV/astrometric solution is 1.3. The derived physical parameters of the 
primary and secondary are shown in Table~5. Their errors include systematic
contribution from the assumed brightness ratio (10 percent in $r$)
and the parallax ($0.85$ in $21.8$ mas). The masses of the components
are $1.38\pm0.13 M_{\odot}$ and $1.39\pm0.13 M_{\odot}$. Unfortunately, 
we cannot independently determine the distance to Hyades using
the current set of $V^2$ measurements for HD~27483. The
star is a perfect candidate for future measurements with CHARA (that
will offer a much longer baseline) since for a 3-day period binary it is
easy to acquire a complete orbital phase coverage in a relatively
short period of time. However, our mass determination of the components
and the Hipparcos distance (corresponding to $M_V=3.6$ mag for HD~27483~A and B 
if $r=1$) place the stars in the mass-luminosity diagram very close
to the prediction based on current theoretical isochrones for
the Hyades cluster \cite[see Fig.~4 from][]{Leb:01::}.

\section{Summary}

We have resolved two double-lined spectroscopic binaries, HD~6118 and HD~27483,
with the Palomar Testbed Interferometer. The data collected with PTI in 2002-2003 
in the K band allow us to determine astrometric orbits and when combined with 
the radial velocity measurements also to derive all physical parameters of the 
systems. The masses of the components of HD 6118 are $2.65\pm0.27 M_{\odot}$ and 
$2.36\pm0.24 M_{\odot}$ and the distance to the system is $112.9\pm0.9$ pc.
Using the theoretical isochrones from \cite{Ber:94::}, we determine the age
of HD~6118~AB as approximately 160-200 Myr. HD~27483 is a double-lined 
spectroscopic binary in the Hyades open cluster. The masses of its components
are $1.38\pm0.13 M_{\odot}$ and $1.39\pm0.13 M_{\odot}$. Unfortunately, the
system is only partly resolved with PTI and hence we are unable to reliably 
determine its apparent semi-major axis and thus the distance. However, our
measurement of the component masses and the distance to the star from the 
Hipparcos catalog place the binary in the mass-luminosity diagram
very close to the theoretical prediction by the current models for 
the Hyades cluster \citep{Leb:01::}.

\acknowledgements

The data presented in this paper were obtained at the Palomar
Observatory using the Palomar Testbed Interferometer, which is supported by
NASA contracts to the Jet Propulsion Laboratory. Science operations with PTI
are possible through the efforts of the PTI Collaboration and excellent
observational work of Kevin Rykoski. This research has made use of 
software produced at the Michelson Science Center,  
California Institute of Technology and the SIMBAD database,   
operated at CDS, Strasbourg, France. M.K. gratefully acknowledges the
support of NASA through the Michelson fellowship program.

\clearpage

%
%

\figcaption[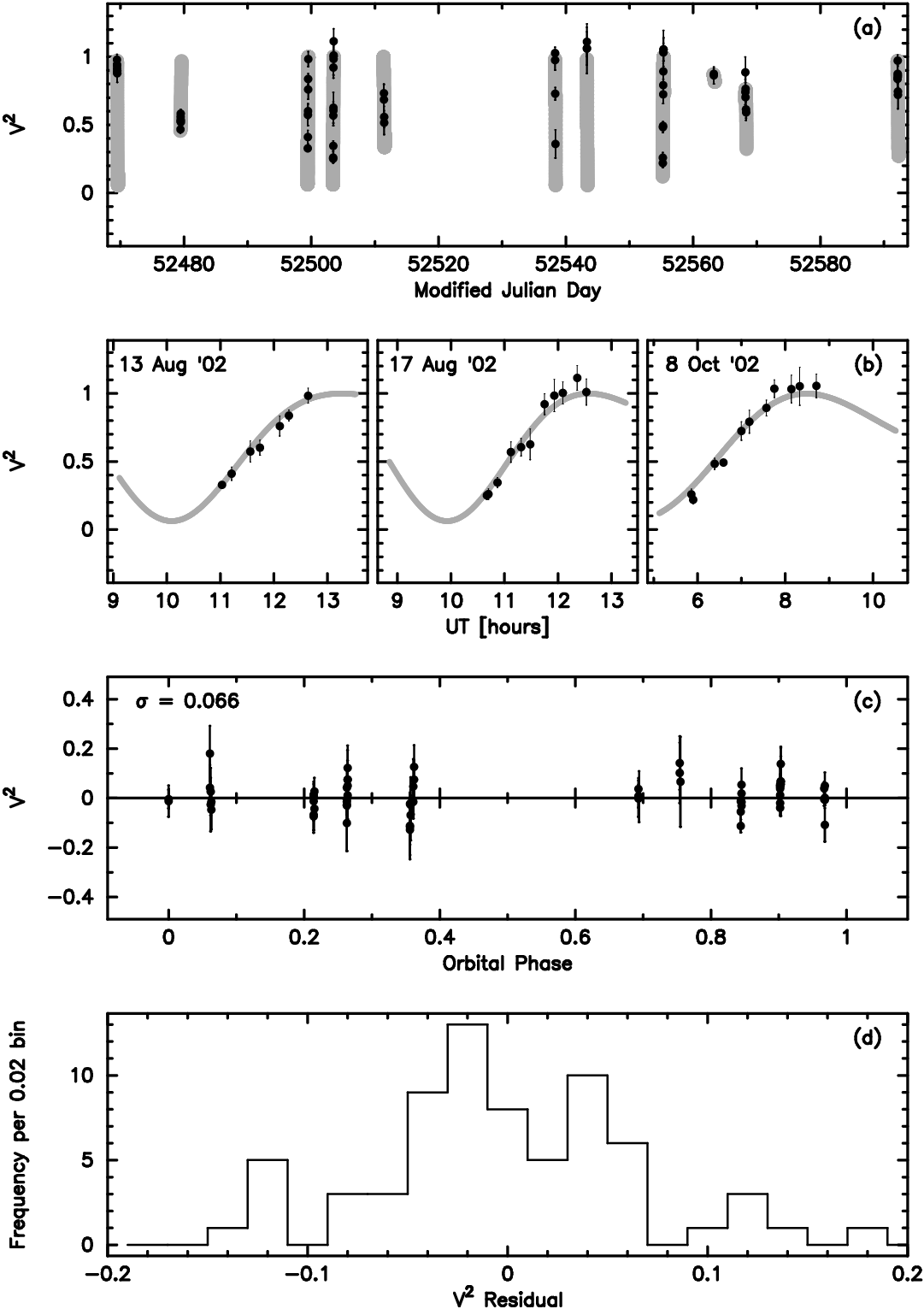]{({\it a,b}) Observed (filled circles) and modeled (gray solid line) 
interferometric visibilities ($V^2$) of HD~6118 as a function of time. Blurred gray lines
in the top panel correspond to the $V^2$ evolution throughout each observing night.
Details of representative examples of the $V^2$ variations are shown in panel {\it (b)}.
({\it c,d}) Best-fit residuals and histogram of $V^2$.}

\figcaption[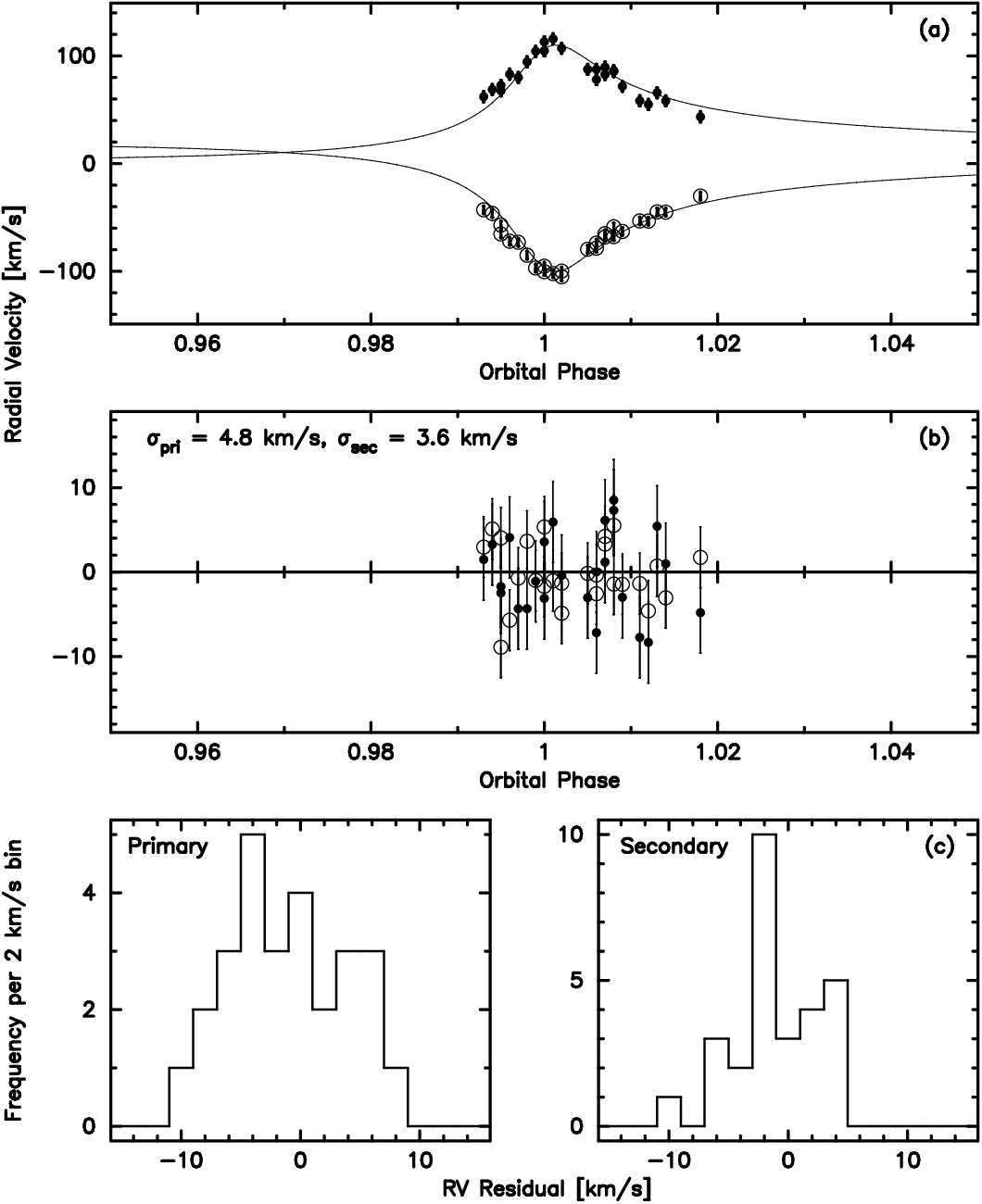]{({\it a}) Observed (filled circles for the primary and open circles
for the secondary) and modeled (solid line) radial velocities (RV) of HD~6118. 
({\it b,c}) Best-fit residuals as a function of time and histogram of the RVs.}

\figcaption[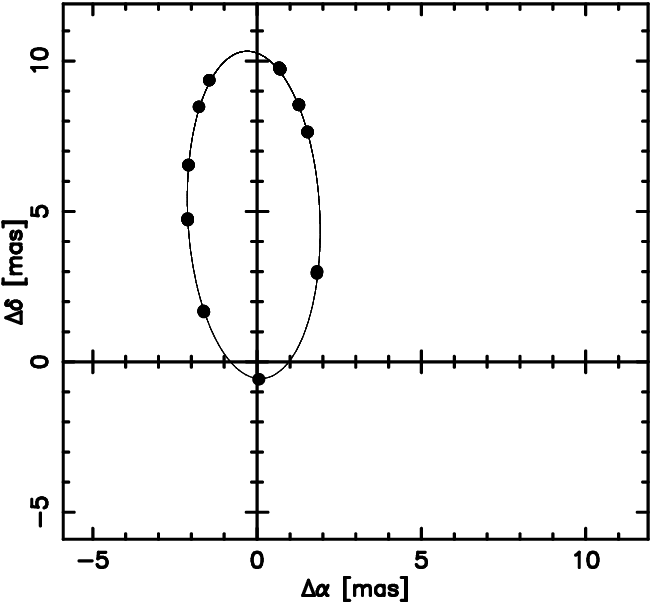]{The apparent orbit of the secondary with respect to the 
primary of HD~6118 from the best-fit V2 model. Note that the filled circles reflect 
the orbital phase coverage of $V^2$ measurements and not the angular separation 
measurements.}

\figcaption[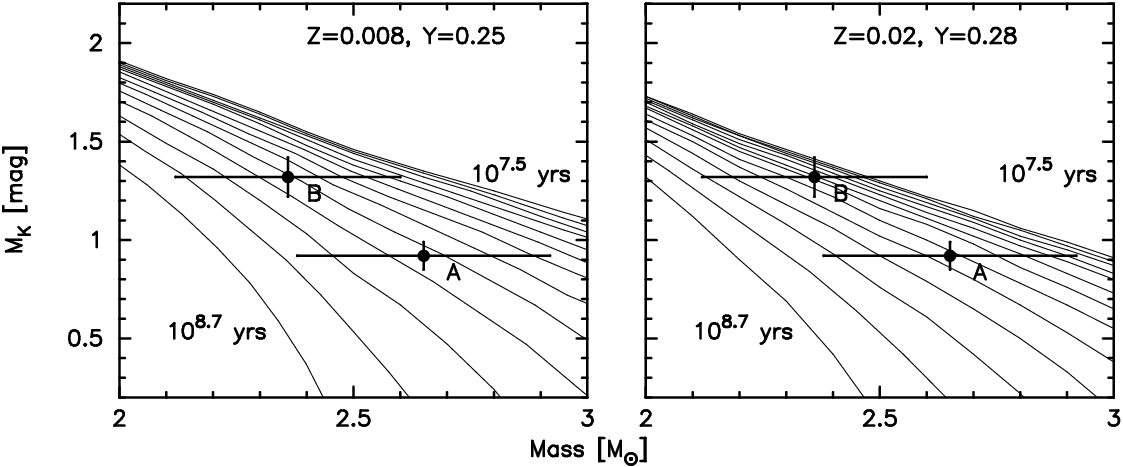]{The primary (A) and secondary (B) of HD~6118 in the 
mass-absolute magnitude diagram. The theoretical isochrones from \cite{Ber:94::} 
(denoted with solid lines) are for $10^{7.5}, 10^{7.6}, ... , 10^{8.6}, 10^{8.7}$ years.}

\figcaption[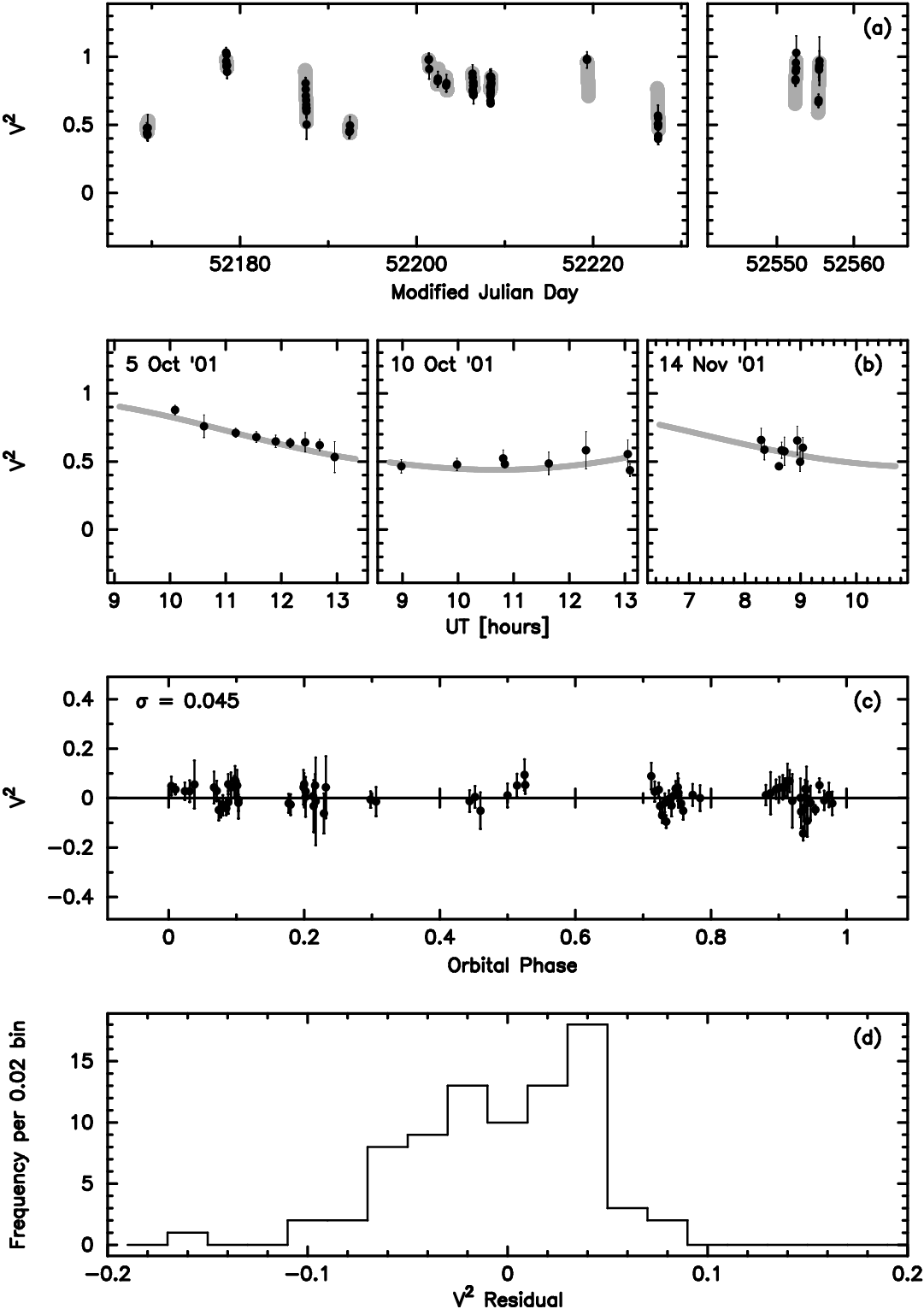]{({\it a,b}) Observed (filled circles) and modeled (gray solid line) 
interferometric visibilities ($V^2$) of HD~27483 as a function of time. 
Blurred gray lines in the top panel correspond to the $V^2$ evolution throughout each 
observing night. Details of representative examples of the $V^2$ variations are shown 
in panel {\it (b)}. ({\it c,d}) Best-fit residuals and histogram of $V^2$.}

\figcaption[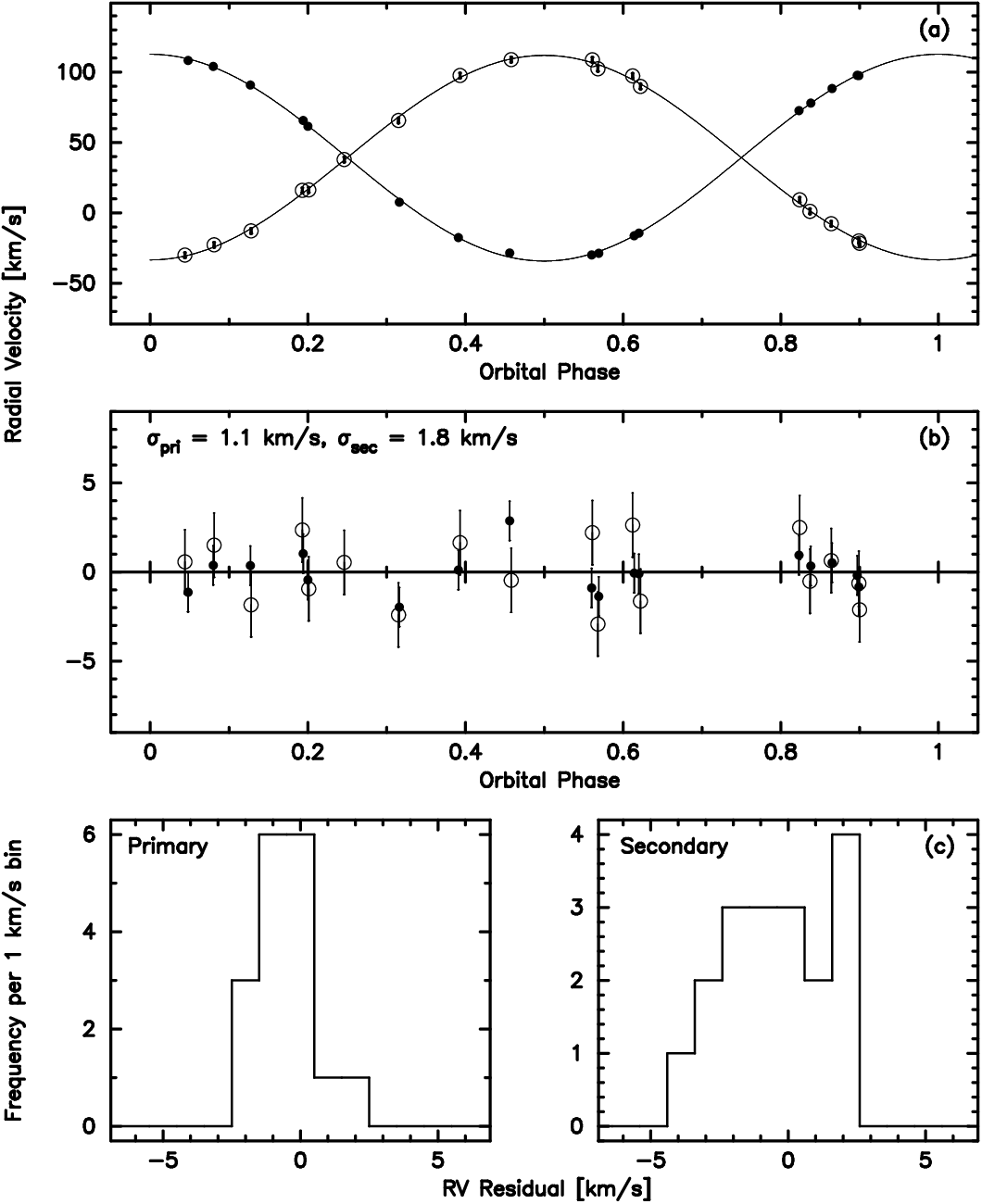]{({\it a}) Observed (filled circles for the primary and open circles
for the secondary) and modeled (solid line) radial velocities (RV) of HD~27483. 
({\it b,c}) Best-fit residuals as a function of time and histogram of the RVs.}

\figcaption[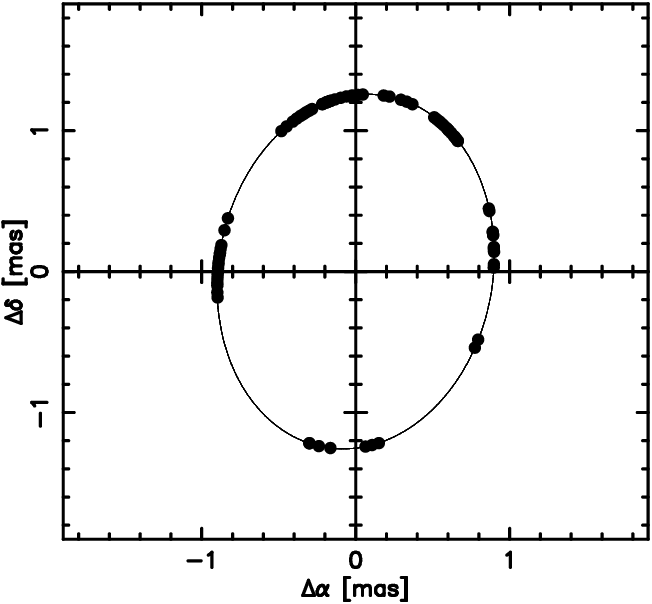]{The apparent orbit of the secondary with respect to the 
primary of HD~27483 from the best-fit V2 model. Note that the filled circles reflect 
the orbital phase coverage of $V^2$ measurements and not the angular separation 
measurements.}

\clearpage

%
%

%
%

\begin{figure}
\figurenum{1}
\epsscale{0.8}
\plotone{f1.eps}
\caption{}
\end{figure}

%
%

\begin{figure}
\figurenum{2}
\epsscale{0.8}
\plotone{f2.eps}
\caption{}
\end{figure}

%
%

\begin{figure}
\figurenum{3}
\epsscale{0.8}
\plotone{f3.eps}
\caption{}
\end{figure}

%
%

\begin{figure}
\figurenum{4}
\epsscale{1.0}
\plotone{f4.eps}
\caption{}
\end{figure}

%
%

\begin{figure}
\figurenum{5}
\epsscale{0.8}
\plotone{f5.eps}
\caption{}
\end{figure}

%
%

\begin{figure}
\figurenum{6}
\epsscale{0.8}
\plotone{f6.eps}
\caption{}
\end{figure}

%
%

\begin{figure}
\figurenum{7}
\epsscale{0.8}
\plotone{f7.eps}
\caption{}
\end{figure}

\clearpage
 
%
%

%
%

\begin{deluxetable}{llcccc}
\tablewidth{495pt}
\tablecaption{Calibration stars for HD~6118 and HD~27483\tablenotemark{a} }
\tablehead{\colhead{Target} & \colhead{Calibrator} & \colhead{Spectral} &  
\colhead{Magnitude} & \colhead{Angular Separation} & \colhead{Adopted} \\
\colhead{} & \colhead{} & \colhead{Type} &  \colhead{}
& \colhead{from Target (deg)} & \colhead{Diameter (mas)} }
\startdata
HD~6118 & & B9.5V+B9.5V & 5.5 V, 5.6 K & & \\
 & HD~7034 & F0V & 5.2 V, 4.5 K & 1.8 & 0.52$\pm$0.1 \\
 & HD~8673 & F7V & 6.3 V, 5.0 K & 5.6 & 0.38$\pm$0.1 \\
 & HD~11007 & F8V & 5.8 V, 4.4 K & 9.7 & 0.45$\pm$0.1 \\
 \hline
HD~27483 & & F6V+F6V & 6.2 V, 5.1 K& & \\
 & HD~21686 & A0V & 5.1 V, 5.1 K & 12.6 & 0.32$\pm$0.09 \\
 & HD~24357 & A4V & 6.0 V, 5.0 K & 7.5 & 0.40$\pm$0.09 \\
\enddata
\tablenotetext{a}{The adopted diameters of the calibrators are determined 
from their effective temperature and bolometric flux derived from archival 
photometry.}
\end{deluxetable}

%
%

\begin{deluxetable}{lc}
\tablewidth{435pt}
\tablecaption{Best-fit Orbital Parameters for HD~6118\tablenotemark{a}}
\tablehead{\colhead{Parameter} & \colhead{HD~6118} }
\startdata
Apparent semi-major axis, $\hat{a}$ (mas) \dotfill & 5.56 $\pm$ 0.04(0.03/0.03)\\
Period, $P$ (d) \dotfill &  81.12625 $\pm$ 2.7(2.0/1.7)$\times 10^{-4}$\\
Time of periastron, $T_p$ (MJD) \dotfill & 31308.153 $\pm$ 0.023(0.021/0.009)\\
Eccentricity, $e$ \dotfill &  0.8956 $\pm$ 0.0020(0.0018/0.0008)\\
Longitude of the periastron, $\omega$ (deg) \dotfill & 346.6 $\pm$ 2.0(1.8/0.9)\\
Longitude of the ascending node, $\Omega$ (deg) \dotfill &  167.8 $\pm$ 1.7(1.5/0.7)\\
Inclination, $i$ (deg) \dotfill & 143.4 $\pm$ 1.3(1.0/0.7) \\
Brightness ratio (K band), $r_K$ \dotfill & 0.69 $\pm$ 0.10(0.08/0.06)\\
Velocity amplitude of the primary, $K_1$ (km/s) \dotfill & 53.2 $\pm$ 1.9(1.9/0.3)\\
Velocity amplitude of the secondary, $K_2$ (km/s) \dotfill & 59.6 $\pm$ 1.6(1.6/0.1)\\
Systemic velocity, $\gamma$ (km/s) \dotfill &  10.5 $\pm$ 2.3(2.3/0.2)\\
Reduced $\chi^2$, $\chi^2/DOF$ \dotfill & 1.06\\
\enddata
\tablenotetext{a}{Figures in parentheses are the $1\sigma$ statistical and systematic
uncertainties contributing to the total error of each parameter.}
\end{deluxetable}

%
%

\begin{deluxetable}{lcc}
\tabletypesize{\small}
\tablewidth{490pt}
\tablecaption{Physical Parameters for HD~6118\tablenotemark{a}}
\tablehead{\colhead{Parameter} & \colhead{Primary} &  \colhead{Secondary}}
\startdata
Semi-major axis, $a_{1,2}$ (AU)\dotfill  & 0.296 $\pm$ 0.011(0.011/0.002) 
 & 0.332 $\pm$ 0.010(0.009/0.001) \\
Mass, $M$ (M$_{\odot}$)\dotfill   &  2.65 $\pm$ 0.27(0.23/0.16)
 & 2.36 $\pm$ 0.24(0.21/0.12) \\
Parallax, $\kappa$ (mas)\dotfill &  & \hspace{-5cm}{\mbox{$8.86 \pm 0.07(0.05/0.04)$}} \\
Distance, $d$ (pc)\dotfill & & \hspace{-5cm}{\mbox{$112.9 \pm 0.9$}} \\
Absolute K magnitude, $M_{K}^{2MASS}$ (mag) \dotfill & 0.92 $\pm$ 0.07 & 1.32 $\pm$ 0.10\\
Spectral type \dotfill  & B9.5V & B9.5V \\
Diameter, $\theta$ (mas)\dotfill & 0.16 & 0.15 \\
\enddata
\tablenotetext{a}{Figures in parentheses are the $1\sigma$ statistical and systematic
uncertainties contributing to the total error of each parameter.}
\end{deluxetable}

%
%

\begin{deluxetable}{rrrrrrrr}
\tablewidth{430pt}
\tablecaption{The K-band data set for HD~6118}
\tablehead{
\colhead{MJD} & \colhead{$V^2$} & \colhead{$\sigma_{V^2}$} & 
\colhead{$(O-C)$} & \colhead{$\lambda$} & 
\colhead{$u$} & \colhead{$v$} & \colhead{Orbital} \\
\colhead{} & \colhead{} & \colhead{} & 
\colhead{($V^2$)} & \colhead{($\mu$m)} & 
\colhead{(m)} & \colhead{(m)} & \colhead{Phase}
}
\startdata
52469.4716&  0.885&  0.026& -0.111& 2.206&  -62.29763&  -88.76658&  0.844\\
52469.4750&  0.943&  0.044& -0.054& 2.212&  -61.59159&  -89.46617&  0.844\\
52469.4822&  0.978&  0.040& -0.012& 2.216&  -60.01158&  -90.91343&  0.844\\
52469.4896&  0.937&  0.024& -0.029& 2.213&  -58.23851&  -92.37724&  0.845\\
52469.4971&  0.913&  0.044& -0.015& 2.213&  -56.33916&  -93.79407&  0.845\\
52469.5043&  0.896&  0.033&  0.020& 2.209&  -54.38969&  -95.11475&  0.845\\
52469.5114&  0.879&  0.066&  0.055& 2.215&  -52.35453&  -96.37247&  0.845\\
52479.4491&  0.467&  0.026& -0.005& 2.209&  -80.25213&   -6.10222&  0.967\\
52479.4593&  0.528&  0.027&  0.039& 2.210&  -81.60143&   -8.83451&  0.967\\
52479.4694&  0.559&  0.054&  0.049& 2.208&  -82.60869&  -11.58149&  0.968\\
52479.4829&  0.537&  0.014& -0.007& 2.212&  -83.43723&  -15.30779&  0.968\\
52479.4964&  0.583&  0.031&  0.001& 2.208&  -83.66175&  -19.07214&  0.968\\
52479.5097&  0.520&  0.068& -0.108& 2.215&  -83.28990&  -22.75824&  0.968\\
52499.4597&  0.328&  0.016&  0.013& 2.212&  -83.02389&  -24.03339&  0.214\\
52499.4669&  0.410&  0.049&  0.008& 2.215&  -82.46938&  -26.01122&  0.214\\
52499.4818&  0.573&  0.078& -0.012& 2.218&  -80.79138&  -30.03950&  0.214\\
52499.4894&  0.601&  0.056& -0.073& 2.217&  -79.65878&  -32.06352&  0.214\\
52499.5044&  0.760&  0.073& -0.066& 2.213&  -76.87996&  -35.97718&  0.214\\
52499.5115&  0.837&  0.041& -0.042& 2.214&  -75.33631&  -37.76197&  0.215\\
52499.5265&  0.983&  0.055&  0.028& 2.219&  -71.56876&  -41.42155&  0.215\\
52503.4446&  0.251&  0.031& -0.007& 2.213&  -83.27129&  -22.85984&  0.263\\
52503.4459&  0.260&  0.039& -0.015& 2.212&  -83.20231&  -23.21733&  0.263\\
52503.4530&  0.345&  0.037& -0.030& 2.211&  -82.72204&  -25.18546&  0.263\\
52503.4635&  0.569&  0.075&  0.042& 2.212&  -81.70878&  -28.06198&  0.263\\
52503.4712&  0.605&  0.064& -0.027& 2.214&  -80.74277&  -30.13430&  0.263\\
52503.4783&  0.626&  0.113& -0.100& 2.213&  -79.68035&  -32.02812&  0.263\\
52503.4898&  0.921&  0.077&  0.076& 2.219&  -77.63532&  -35.01885&  0.264\\
52503.4970&  0.984&  0.119&  0.076& 2.215&  -76.14014&  -36.85894&  0.264\\
52503.5037&  1.004&  0.081&  0.051& 2.211&  -74.60828&  -38.53705&  0.264\\
52503.5149&  1.114&  0.091&  0.123& 2.213&  -71.73623&  -41.27522&  0.264\\
52503.5219&  1.010&  0.096&  0.012& 2.223&  -69.77414&  -42.91569&  0.264\\
52511.4097&  0.733&  0.067& -0.015& 2.225&  -48.33889&  -98.55608&  0.361\\
52511.4251&  0.685&  0.111&  0.047& 2.228&  -43.21250& -100.88707&  0.361\\
52511.4474&  0.558&  0.078&  0.074& 2.227&  -35.01492& -103.79922&  0.362\\
52511.4621&  0.516&  0.088&  0.125& 2.221&  -29.24109& -105.36906&  0.362\\
52538.3073&  1.027&  0.044&  0.038& 2.218&  -82.52664&  -11.31196&  0.693\\
52538.3217&  0.976&  0.074& -0.001& 2.217&  -83.43361&  -15.27362&  0.693\\
52538.3450&  0.729&  0.046&  0.008& 2.214&  -83.44924&  -21.76152&  0.693\\
52538.3663&  0.360&  0.103&  0.004& 2.218&  -81.89032&  -27.61842&  0.694\\
52543.3030&  1.062&  0.122&  0.103& 2.224&  -83.19274&  -13.88360&  0.754\\
52543.3044&  1.110&  0.108&  0.143& 2.222&  -83.26775&  -14.26941&  0.754\\
52543.3201&  1.060&  0.182&  0.067& 2.224&  -83.66703&  -18.61957&  0.755\\
52555.2441&  0.259&  0.038&  0.012& 2.226&  -60.75089&  -90.25274&  0.902\\
52555.2458&  0.219&  0.032& -0.039& 2.229&  -60.36032&  -90.60454&  0.902\\
52555.2663&  0.483&  0.041&  0.046& 2.225&  -55.25302&  -94.54373&  0.902\\
52555.2746&  0.492&  0.021& -0.020& 2.227&  -52.93196&  -96.02605&  0.902\\
52555.2915&  0.724&  0.068&  0.055& 2.224&  -47.74311&  -98.85137&  0.902\\
52555.2994&  0.792&  0.084&  0.051& 2.218&  -45.12403& -100.07088&  0.902\\
52555.3155&  0.893&  0.058&  0.040& 2.223&  -39.45833& -102.33171&  0.902\\
52555.3230&  1.035&  0.065&  0.140& 2.225&  -36.67357& -103.28025&  0.903\\
52555.3393&  1.032&  0.102&  0.069& 2.228&  -30.31187& -105.10536&  0.903\\
52555.3470&  1.053&  0.140&  0.070& 2.227&  -27.23835& -105.83424&  0.903\\
52555.3630&  1.055&  0.087&  0.058& 2.230&  -20.55383& -107.11264&  0.903\\
52563.2452&  0.875&  0.038& -0.003& 2.225&  -55.07166&  -94.66790&  0.000\\
52563.2630&  0.863&  0.063& -0.011& 2.223&  -49.85921&  -97.77326&  0.000\\
52568.2017&  0.741&  0.092&  0.043& 2.235&  -62.23040&  -88.83121&  0.061\\
52568.2086&  0.886&  0.112&  0.180& 2.238&  -60.76237&  -90.24129&  0.061\\
52568.2242&  0.703&  0.108& -0.025& 2.235&  -57.00814&  -93.30844&  0.062\\
52568.2583&  0.765&  0.096&  0.027& 2.232&  -46.96657&  -99.22493&  0.062\\
52568.3072&  0.614&  0.080& -0.047& 2.240&  -28.98546& -105.43060&  0.063\\
52568.3265&  0.591&  0.041& -0.015& 2.240&  -20.99243& -107.04018&  0.063\\
52592.1255&  0.843&  0.135& -0.110& 2.228&  -64.26670&  -86.59406&  0.356\\
52592.1271&  0.837&  0.102& -0.126& 2.225&  -63.98127&  -86.93272&  0.356\\
52592.1352&  0.875&  0.072& -0.115& 2.228&  -62.41920&  -88.64280&  0.356\\
52592.1472&  0.972&  0.041& -0.021& 2.228&  -59.81671&  -91.08260&  0.356\\
52592.1724&  0.866&  0.107& -0.023& 2.230&  -53.25580&  -95.83075&  0.357\\
52592.1741&  0.854&  0.032& -0.020& 2.228&  -52.77823&  -96.12117&  0.357\\
52592.1856&  0.720&  0.102& -0.069& 2.227&  -49.29094&  -98.07232&  0.357\\
52592.1877&  0.745&  0.031& -0.028& 2.227&  -48.61715&  -98.41827&  0.357\\
\enddata                                                            
\end{deluxetable}                                                   

%
%

\begin{deluxetable}{lc}
\tablewidth{470pt}
\tablecaption{Best-fit Orbital Parameters for HD~27483\tablenotemark{a}}
\tablehead{\colhead{Parameter} & \colhead{HD~27483} }
\startdata
Period, $P$ (d) \dotfill & 3.0591080 $\pm$ 1.1(1.0/0.5)$\times 10^{-5}$\\
Time of periastron, $T_p$ (MJD) \dotfill & 44497.185696 $\pm$ 0.0026(0.0026/0.0006)\\
Eccentricity (assumed), $e$ \dotfill &  0.0\\
Longitude of the ascending node, $\Omega$ (deg) \dotfill & 7.3 $\pm$ 3.6(3.2/1.7)\\
Inclination, $i$ (deg) \dotfill & 45.1 $\pm$ 1.7(0.6/1.6) \\
Brightness ratio (K band, assumed), $r_K$ \dotfill & 1.0\\
Velocity amplitude of the primary, $K_1$ (km/s) \dotfill & 73.4 $\pm$ 0.4(0.4/0.03)\\
Velocity amplitude of the secondary, $K_2$ (km/s) \dotfill & 72.5 $\pm$ 0.6(0.6/0.07)\\
Systemic velocity, $\gamma$ (km/s) \dotfill &  39.2 $\pm$ 0.2(0.2/0.007)\\
{\it Apparent semi-major axis}, $\hat{a}$ (mas) \dotfill & 1.26 $\pm$ 0.05(0.05/0.0007)\\
Reduced $\chi^2$, $\chi^2/DOF$ \dotfill & 1.3 \\
\enddata
\tablenotetext{a}{Figures in parentheses are the $1\sigma$ statistical and systematic
uncertainties contributing to the total error of each parameter.}
\end{deluxetable}

%
%

\begin{deluxetable}{lcc}
\tabletypesize{\small}
\tablewidth{500pt}
\tablecaption{Physical Parameters for HD~27483\tablenotemark{a}}
\tablehead{\colhead{Parameter} & \colhead{Primary} &  \colhead{Secondary}}
\startdata
Semi-major axis, $a_{1,2}$ (AU)\dotfill  & 0.02915 $\pm$ 1.4(1.4/0.1)$\times 10^{-4}$ 
 &  0.02878 $\pm$ 2.4(2.4/0.3)$\times 10^{-4}$ \\
Mass, $M$ (M$_{\odot}$)\dotfill   &  1.38 $\pm$ 0.13(0.05/0.12)
 &  $\pm$ 1.39 0.13(0.05/0.12) \\
Parallax (from Hipparcos), $\kappa$ (mas)\dotfill &  & \hspace{-6cm}{$21.8 \pm 0.85 $} \\
Distance (from Hipparcos), $d$ (pc)\dotfill & & \hspace{-6cm}{$45.9 \pm 1.8 $} \\
Absolute K magnitude, $M_{K}^{2MASS}$ (mag) \dotfill & 2.51 $\pm$ 0.07 & 2.51 $\pm$ 0.07 \\
Spectral type \dotfill  & F6V &  F6V \\
Diameter, $\theta$ (mas)\dotfill & 0.25 & 0.25 \\
\enddata
\tablenotetext{a}{Figures in parentheses are the $1\sigma$ statistical and systematic
uncertainties contributing to the total error of each parameter.}
\end{deluxetable}

%
%

\begin{deluxetable}{rrrrrrrr}
\tablewidth{430pt}
\tablecaption{The K-band data set for HD~27483}
\tablehead{
\colhead{MJD} & \colhead{$V^2$} & \colhead{$\sigma_{V^2}$} & 
\colhead{$(O-C)$} & \colhead{$\lambda$} & 
\colhead{$u$} & \colhead{$v$} & \colhead{Orbital} \\
\colhead{} & \colhead{} & \colhead{} & 
\colhead{($V^2$)} & \colhead{($\mu$m)} & 
\colhead{(m)} & \colhead{(m)} & \colhead{Phase}
}
\startdata
52169.4411&  0.477&  0.038&  0.049& 2.216&  -60.04103&  -90.67138&  0.004\\
52169.4617&  0.442&  0.020&  0.034& 2.211&  -54.81478&  -92.46702&  0.011\\
52169.5040&  0.425&  0.035&  0.028& 2.217&  -41.34146&  -95.55666&  0.025\\
52169.5256&  0.432&  0.044&  0.028& 2.214&  -33.27606&  -96.77422&  0.032\\
52169.5473&  0.478&  0.097&  0.055& 2.219&  -24.54295&  -97.72336&  0.039\\
52178.4216&  1.031&  0.037&  0.038& 2.206&  -81.67179&  -16.86810&  0.940\\                                                 
52178.4457&  0.966&  0.070& -0.020& 2.209&  -83.47920&  -19.87857&  0.948\\                                                 
52178.4663&  0.930&  0.018& -0.046& 2.211&  -83.50226&  -22.47998&  0.954\\
52178.4863&  1.015&  0.028&  0.053& 2.212&  -82.18878&  -24.97510&  0.961\\
52178.5066&  0.936&  0.039& -0.009& 2.209&  -79.51184&  -27.46343&  0.968\\
52178.5269&  0.939&  0.049&  0.013& 2.209&  -75.54893&  -29.83948&  0.974\\
52178.5439&  0.889&  0.047& -0.020& 2.208&  -71.25532&  -31.73276&  0.980\\
52187.4202&  0.806&  0.041&  0.013& 2.214&  -52.64961&  -93.08626&  0.881\\
52187.4419&  0.761&  0.085&  0.021& 2.219&  -45.84172&  -94.70129&  0.888\\
52187.4660&  0.714&  0.039&  0.037& 2.220&  -37.26240&  -96.21923&  0.896\\
52187.4814&  0.681&  0.044&  0.043& 2.218&  -31.34041&  -97.01585&  0.901\\
52187.4957&  0.647&  0.049&  0.044& 2.216&  -25.58201&  -97.62907&  0.906\\
52187.5065&  0.622&  0.036&  0.044& 2.217&  -21.05021&  -98.01201&  0.910\\
52187.5179&  0.623&  0.070&  0.069& 2.219&  -16.19633&  -98.33250&  0.913\\
52187.5288&  0.601&  0.044&  0.069& 2.218&  -11.47321&  -98.56032&  0.917\\
52187.5401&  0.503&  0.108& -0.010& 2.223&   -6.51337&  -98.71406&  0.920\\
52192.3743&  0.448&  0.048&  0.010& 2.222&  -60.93263&  -90.30497&  0.501\\
52192.4156&  0.456&  0.047&  0.051& 2.222&  -49.91011&  -93.78621&  0.514\\
52192.4503&  0.496&  0.063&  0.094& 2.224&  -38.05173&  -96.09529&  0.526\\
52192.4517&  0.456&  0.035&  0.053& 2.224&  -37.50915&  -96.17839&  0.526\\
52201.3806&  0.977&  0.044& -0.011& 2.214&  -83.38707&  -19.58007&  0.445\\
52201.4042&  0.983&  0.044&  0.006& 2.211&  -83.48201&  -22.55729&  0.453\\
52201.4276&  0.911&  0.074& -0.050& 2.215&  -81.75339&  -25.48721&  0.460\\
52202.3848&  0.820&  0.044&  0.014& 2.212&  -83.60642&  -20.45829&  0.773\\
52202.4193&  0.840&  0.052&  0.000& 2.213&  -82.32174&  -24.80380&  0.784\\
52203.3499&  0.789&  0.046& -0.017& 2.215&  -81.25220&  -16.43679&  0.089\\
52203.3879&  0.806&  0.063&  0.051& 2.215&  -83.66799&  -21.18956&  0.101\\
52206.3447&  0.877&  0.064&  0.042& 2.214&  -81.61656&  -16.80890&  0.068\\
52206.3551&  0.848&  0.041&  0.029& 2.220&  -82.64579&  -18.09778&  0.071\\
52206.3660&  0.755&  0.042& -0.048& 2.216&  -83.34320&  -19.45886&  0.075\\
52206.3740&  0.747&  0.033& -0.046& 2.218&  -83.60796&  -20.46804&  0.077\\
52206.3844&  0.750&  0.041& -0.030& 2.217&  -83.63433&  -21.78276&  0.081\\
52206.3995&  0.723&  0.028& -0.042& 2.223&  -83.03426&  -23.68374&  0.085\\
52206.4086&  0.812&  0.041&  0.056& 2.220&  -82.30747&  -24.82260&  0.088\\
52206.4217&  0.790&  0.061&  0.043& 2.221&  -80.78114&  -26.44374&  0.093\\
52206.4352&  0.787&  0.043&  0.047& 2.216&  -78.64555&  -28.06514&  0.097\\
52206.4404&  0.811&  0.057&  0.072& 2.216&  -77.67525&  -28.67474&  0.099\\
52206.4541&  0.731&  0.076& -0.008& 2.222&  -74.68480&  -30.26203&  0.103\\
52206.4556&  0.718&  0.033& -0.020& 2.217&  -74.33879&  -30.42443&  0.104\\
52208.3170&  0.854&  0.054&  0.089& 2.215&  -78.24352&  -14.11855&  0.712\\
52208.3315&  0.779&  0.041&  0.026& 2.214&  -80.61362&  -15.85313&  0.717\\
52208.3500&  0.779&  0.028&  0.034& 2.212&  -82.67478&  -18.14216&  0.723\\
52208.3576&  0.712&  0.028& -0.032& 2.213&  -83.19229&  -19.08823&  0.726\\
52208.3679&  0.676&  0.030& -0.070& 2.219&  -83.59498&  -20.38941&  0.729\\
52208.3772&  0.730&  0.038& -0.018& 2.214&  -83.65442&  -21.57058&  0.732\\
52208.3847&  0.657&  0.025& -0.095& 2.213&  -83.49330&  -22.51368&  0.734\\
52208.3982&  0.759&  0.035& -0.004& 2.217&  -82.73269&  -24.21070&  0.739\\
52208.4079&  0.742&  0.044& -0.030& 2.214&  -81.81728&  -25.41565&  0.742\\
52208.4150&  0.788&  0.028&  0.008& 2.215&  -80.95680&  -26.28502&  0.744\\
52208.4261&  0.834&  0.030&  0.039& 2.216&  -79.28248&  -27.62896&  0.748\\
52208.4362&  0.855&  0.056&  0.044& 2.219&  -77.41629&  -28.82800&  0.751\\
52208.4398&  0.838&  0.059&  0.022& 2.219&  -76.68183&  -29.24409&  0.752\\
52208.4507&  0.811&  0.032& -0.022& 2.212&  -74.20814&  -30.48483&  0.756\\
52208.4615&  0.803&  0.036& -0.051& 2.220&  -71.41482&  -31.67054&  0.760\\
52219.2873&  0.984&  0.033& -0.004& 2.214&  -63.61692&  -89.06337&  0.298\\
52219.3115&  0.977&  0.059& -0.013& 2.213&  -58.40521&  -91.28894&  0.306\\
52227.3454&  0.568&  0.078&  0.002& 2.231&  -41.44225&  -95.53847&  0.933\\
52227.3478&  0.506&  0.067& -0.055& 2.227&  -40.59268&  -95.68519&  0.933\\
52227.3588&  0.397&  0.028& -0.143& 2.229&  -36.50589&  -96.32982&  0.937\\
52227.3610&  0.496&  0.054& -0.039& 2.227&  -35.68126&  -96.44839&  0.938\\
52227.3631&  0.487&  0.090& -0.045& 2.228&  -34.89705&  -96.55776&  0.938\\
52227.3726&  0.550&  0.092&  0.036& 2.227&  -31.15997&  -97.03593&  0.941\\
52227.3746&  0.419&  0.064& -0.092& 2.227&  -30.36469&  -97.12892&  0.942\\
52227.3766&  0.506&  0.070& -0.002& 2.228&  -29.58014&  -97.21776&  0.943\\
52552.4301&  0.834&  0.045&  0.056& 2.221&  -49.87012&  -93.79909&  0.200\\
52552.4378&  0.826&  0.042&  0.027& 2.222&  -47.43702&  -94.36073&  0.203\\
52552.4707&  0.894&  0.044&  0.009& 2.223&  -35.76859&  -96.43770&  0.214\\
52552.4788&  0.956&  0.046&  0.052& 2.221&  -32.62244&  -96.85885&  0.216\\
52552.5181&  0.913&  0.081& -0.061& 2.228&  -16.42641&  -98.32007&  0.229\\
52552.5273&  1.029&  0.125&  0.045& 2.226&  -12.45217&  -98.52052&  0.232\\
52555.4198&  0.667&  0.038& -0.021& 2.227&  -50.51831&  -93.64104&  0.178\\
52555.4276&  0.683&  0.042& -0.026& 2.225&  -48.07954&  -94.21867&  0.180\\
52555.4868&  0.921&  0.070&  0.044& 2.225&  -26.13483&  -97.57822&  0.200\\
52555.4958&  0.905&  0.080&  0.006& 2.225&  -22.41073&  -97.90756&  0.203\\
52555.5321&  0.938&  0.105& -0.031& 2.227&   -6.75279&  -98.71059&  0.214\\
52555.5430&  0.969&  0.177& -0.012& 2.227&   -1.96827&  -98.78206&  0.218\\
\enddata
\end{deluxetable}

\end{document}